\documentclass[twocolumn,showpacs,preprintnumbers,amsmath,amssymb]{revtex4}
\usepackage{graphicx}% Include figure files
\usepackage{dcolumn}% Align table columns on decimal point

\usepackage{bm}% bold math

\newcommand{\ha}{\hat{a}}
\newcommand{\hap}{\hat{a}^+}

\newcommand{\hH}{\hat{H}}

\begin{document}
\title{Zitterbewegung of electrons in graphene in a magnetic field }
\date{\today}
\author{Tomasz M. Rusin*}
\email{Tomasz.Rusin@centertel.pl}
\author{Wlodek Zawadzki\dag}
\affiliation{*PTK Centertel Sp. z o.o., ul. Skierniewicka 10A, 01-230 Warsaw, Poland\\
            \dag Institute of Physics, Polish Academy of Sciences, Al. Lotnik\'ow 32/46, 02-688 Warsaw, Poland}

\begin{abstract}
Electric current and spacial displacement due to trembling motion [Zitterbewegung (ZB)] of electrons
in graphene in the presence of an external magnetic field are described. Contributions of both inequivalent
$K$ points in the Brillouin zone of graphene are considered. It is shown that, when the electrons are
prepared in the form of wave packets, the presence of a quantizing magnetic field $B$ has very
important effects on ZB. (1) For $B\neq 0$ the ZB oscillations are permanent, for $B=0$ they are transient.
(2) For $B\neq 0$ many ZB frequencies appear, for $B=0$ only one frequency is at work. (3) For $B\neq 0$
both interband and intraband (cyclotron) frequencies contribute to ZB, for $B=0$ there are no
intraband frequencies. (4) Magnetic field intensity changes not only the ZB frequencies but the entire
character of ZB spectrum. An emission of electromagnetic dipole radiation by the trembling electrons
is proposed and described. It is argued that graphene in a magnetic field is a promising system for an
experimental observation of Zitterbewegung.
\end{abstract}

\maketitle
\section{Introduction}
The Zitterbewegung (ZB, trembling motion), first described by Schrodinger with the use of the
Dirac equation for free relativistic electrons in a vacuum \cite{Schrodinger30},
has been in recent years a
subject of great theoretical interest since it has been shown that this phenomenon should exist
in many systems in solids \cite{Cannata90,Imry95,Schliemann05,Zawadzki05KP,Zawadzki06,Cserti06,
Katsnelson06,Winkler07,Trauzettel07,Rusin07a,Rusin07b,Zulick07}.
If one deals with two or more interacting energy branches, an
interference of the upper and lower energy states gives rise to the ZB also in the absence of
external fields. A formal similarity between two bands interacting via the $\bf k  \cdot\bf p$ terms in
a solid and the Dirac equation for relativistic electrons in a vacuum allows one to apply
similar theoretical methods for both systems (see \cite{ZawadzkiOPS,ZawadzkiHMF}).
Most of the theoretical ZB treatments
for semiconductors used as a starting point plane electron waves.
However, Lock \cite{Lock79} in his important paper
observed that, since such a wave is not localized, it seems of a limited practicality to speak of
rapid oscillations on the average position of a wave of infinite extent. Using the Dirac equation
Lock demonstrated that, when an electron is represented by a wave packet, the ZB oscillations do not
sustain  their amplitude but become transient. The disappearance of oscillations at large times is
guaranteed by the Riemann-Lebesgues theorem as long as the wave packet is a smoothly varying spatial function.
The transient character of the trembling motion was demonstrated in our recent papers
\cite{Rusin07b,Zawadzki08}.
Since the ZB is by its nature  not a stationary state but a dynamical phenomenon, it is natural to
study it with the use of wave packets (see \cite{Huang52}). These have become of practical use
with the emergence of the femto-second pulse technology \cite{Garraway95}.
Recently, the transient trembling motion was proposed for ultra cold atoms \cite{Clark07,Merkl08}, for photons
in two-dimensional photonic crystals \cite{Zhang08PRL} and for Ramsey interferometry \cite{Bermudez08}.
Most recently, an actual {\it observation} of an acoustic analogue of ZB in
a macroscopic two-dimensional sonic crystal was reported \cite{Zhang08}.

The trembling motion of charge carriers in solids has been described
until present for no external potentials. On the other hand, Lock has remarked that,
when an electron spectrum is discrete, the ZB oscillations do not have to disappear with time.
In the present work we consider the trembling motion of electrons in
solids in the presence of an external magnetic field, see \cite{Rusin07c,Schliemann08}.
The magnetic field is known to cause no
interband electron transitions, so the essential features of ZB, which results from an
interference of positive and negative energy states of the system, are expected not to be destroyed.
On the other hand, introduction of an external field provides an important
parameter affecting the ZB behavior. We show that, indeed, the ZB in
a magnetic field is not damped in time.

We consider the ZB phenomenon in monolayer graphene. This material has recently become an important
subject of investigation in the condensed matter physics and its applications
\cite{Novoselov04,Novoselov05,Geim07}. In particular,
the charge carriers in graphene are considered to represent massless Dirac fermions.
In our approach we look for observable ZB phenomena.  The situation we describe seems to be
the most promising case for an experimental observation of the trembling motion in solids
considered until present.

The paper is organized as follows. First, we consider the Hamiltonian for electrons
in graphene in a magnetic field, its eigenvalues and eigenfunctions. Second, we calculate
carrier velocities and their averages taken over a Gaussian wave packet. Third, we give results for
the ZB of electric current and electron position
and emphasize the essential features introduced by the presence of a magnetic field.
Finally, we propose and describe electromagnetic radiation emitted by the trembling electrons.
The paper is concluded by a summary. In Appendices we discuss additional points related
to the subject.

\section{Preliminaries}
We consider a graphene monolayer in an external magnetic field parallel to the $z$ axis.
As shown in both continuum and tight-binding models, there exist two inequivalent points
$K_1$ and $K_2$ of the Brillouin zone (BZ) of graphene. The Hamiltonian for electrons and
holes at the $K_1$ point is \cite{Wallace47,Slonczewski58}
\begin{equation} \label{H_pi0}
\hH = u\left(\begin{array}{cc}
     0 & \hat{\pi}_x-i\hat{\pi}_y \\  \hat{\pi}_x+i\hat{\pi}_y & 0 \\     \end{array}\right),
\end{equation}
where $u\approx 1\times $10$^8$ cm/s is a characteristic velocity,
$\hat{\bm \pi} =  \hat{\bm p} - q \hat{\bm A}$ is the generalized momentum,
$\hat{\bm A}$ is the vector potential and $q$ is the electron charge.
We remark that the above Hamiltonian is not symmetric in $\hat{p}_x$ and $\hat{p}_y$.
Using the Landau gauge we take $ \hat{\bm A} = (-By,0,0)$, and for an electron
$q=-e$ with $e>0$. Since in the above gauge $\hH$ does not depend on $x$,
we take the wave function in the form
\begin{equation}  \Psi(x,y) = e^{ik_xx}\Phi(y). \end{equation}
Introducing the magnetic radius $L = \sqrt{\hbar/eB}$ and the variable $\xi=y/L-k_xL$, we have
\begin{equation} \label{H_xi}
 \hH = \frac{u\hbar}{L}\left(\begin{array}{cc}
     0 &  -\xi-\partial/\partial \xi \\
      -\xi + \partial/\partial \xi  & 0 \\     \end{array}\right).
\end{equation}
Defining the standard raising and lowering operators for the harmonic oscillator
$\ha= (\xi+ \partial/\partial \xi )/\sqrt{2}$ and $\hap=(\xi -\partial/\partial \xi)/\sqrt{2}$,
the Hamiltonian becomes
\begin{equation} \label{H aap}
 \hH = -\hbar\Omega\left(\begin{array}{cc}  0 & \ha \\ \hap& 0 \\  \end{array}\right),
\end{equation}
where the frequency is $\Omega=\sqrt{2}u/L$.

Next we determine the  eigenstates and eigenenergies of the Hamiltonian $\hH$.
Using a two-component function $\psi=(\psi_1, \psi_2)$, we have
\begin{equation} \label{H_Omega}
 \left\{\begin{array}{rrr} -\hbar\Omega\ha\psi_2&=&E\psi_1, \\ -\hbar\Omega\hap\psi_1&=&E\psi_2. \end{array} \right.
\end{equation}
Expressing $\psi_1$ by $\psi_2$ from the first equation we obtain from the second equation:
 $\hbar^2 \Omega^2\hap\ha\psi_2=E^2\psi_2$. The solution is  $\psi_2=|n\rangle$,
where $|n\rangle$ is the $n$-th state of the harmonic oscillator, and the energy is
\begin{equation} \label{H_Ens}
 E_{ns}=s\hbar\Omega\sqrt{n}.
\end{equation}
Here $n=0,1,\ldots$, and $s=\pm 1$ for the conduction and valence bands, respectively.
Formula (\ref{H_Ens}) was experimentally confirmed for graphene in many works \cite{Novoselov05,Zhang05,Sadowski06}.
The functions $\langle \bm r|n\rangle$ are given by
\begin{equation} \label{H_Hermit}
  \langle \bm r|n\rangle =  \frac{1}{\sqrt{L}}\frac{1}{C_n}e^{-\frac{1}{2}\xi^2}{\rm H}_n(\xi),
\end{equation}
where $C_n=\sqrt{2^nn!\sqrt{\pi}}\ $, and ${\rm H}_n(\xi)$ are the Hermite polynomials.
From Eq. (\ref{H_Omega})  we obtain
\begin{equation} \label{H_psi1}
\psi_1 = -\frac{\hbar\Omega\ha\psi_2}{E} = -\frac{\hbar\Omega \sqrt{n}|n-1\rangle}{s\hbar\Omega\sqrt{n}} =-s|n-1\rangle.
\end{equation}
Each eigenstate $|{\rm n}\rangle$ is labelled by three quantum numbers:
oscillator number $n$, energy branch $s$, and wave vector $k_x$. The complete function is
\begin{equation} \label{H_nskx}
|{\rm n}\rangle \equiv |nk_xs\rangle = \frac{e^{ik_xx}}{\sqrt{4\pi}}
               \left(\begin{array}{c} -s|n-1\rangle \\|n\rangle    \end{array}\right).
\end{equation}
For $n=0$, the first component in Eq. (\ref{H_nskx}) vanishes and the normalization coefficient
is $1/\sqrt{2\pi}$.

\section{Velocities. Zitterbewegung}
We want to calculate the velocity of charge carriers described by a wave packet. In order to
do that we first calculate matrix elements $\langle f|{\rm n}\rangle$ between an arbitrary
two-component function $f=(f^u,f^l)$ and eigenstates (\ref{H_nskx}). A straightforward
manipulation gives
\begin{equation}  \label{V_Matrix_FuFl}
\langle f|{\rm n}\rangle = -sF^u_{n-1} + F^l_{n},
\end{equation}
where
\begin{equation} \label{V_Matrix_Fj}
  F^j_n(k_x) = \frac{1}{\sqrt{2L}C_n} \int g^j(k_x,y)e^{-\frac{1}{2}\xi^2}{\rm H}_{n}(\xi)dy,
\end{equation}
in which
\begin{equation} \label{V_Matrix_gj}
 g^j(k_x,y) = \frac{1}{\sqrt{2\pi}} \int f^j(x,y)e^{ik_xx} dx.
\end{equation}
The superscript $j=u,l$  stands for the upper and lower components of the function $f$.

The Hamilton equations  give the velocity components:
$\hat{v}_i(0)=\partial \hH/\partial \hat{\pi}_i$, with $i=x,y$. We want to
calculate averages of the time-dependent velocity operators $\hat{v}_i(t)$ in the
Heisenberg picture taken on the function $f$. The averages are
\begin{equation} \label{V_vi(t)}
 \bar{v}_i(t) = \sum_{\rm n,n'} e^{iE_{\rm n'}t/\hbar}\langle f|{\rm n'}\rangle \langle {\rm n'}|v_i(0)|\rm n \rangle
                \langle {\rm n}|f\rangle e^{-iE_{\rm n}t/\hbar},
\end{equation}
where the energies and eigenstates
are given in Eqs. (\ref{H_Ens}) and (\ref{H_nskx}). The summation in Eq. (\ref{V_vi(t)}) goes over
all the quantum numbers
\begin{equation} \sum_{\rm n,n'} \rightarrow \int\int dk_x dk_x' \sum_{n,n'}\sum_{s,s'}. \end{equation}

We calculate a contribution to the velocity from the point $K_1$ of the Brillouin zone.
The matrix elements $ \langle {\rm n'}|v_y(0)|{\rm n} \rangle$ and
                    $ \langle {\rm n'}|v_x(0)|{\rm n} \rangle$
can be shown to be
\begin{eqnarray} \label{V_vy_nn}
 \langle {\rm n'}|v_y(0)|{\rm n} \rangle =\frac{iu}{2}\delta_{k_x\!,k_x'} (-s\delta_{n'\!,n-1} + s'\delta_{n'\!,n+1}),
\end{eqnarray}
\begin{equation} \label{V_vx_nn}
 \langle {\rm n'}|v_x(0)|{\rm n} \rangle=-\frac{u}{2}\delta_{k_x\!,k_x'}(s\delta_{n'\!,n-1}+s'\delta_{n'\!,n+1}).
\end{equation}
It is seen that the only non-vanishing matrix elements of the velocity components are those with
the final states $n'=n\pm 1$. Putting the above matrix elements into Eq. (\ref{V_vi(t)}),
we finally obtain after some manipulation for the $K_1$ point of BZ
\begin{widetext}  \begin{eqnarray} \label{V_vy(t)}
\bar{v}_y(t) &=&  u\sum_{n=0}^{\infty} V_n^+\sin(\omega_n^ct) +  u\sum_{n=0}^{\infty}V_n^-\sin(\omega_n^Zt)
                +iu\sum_{n=0}^{\infty} A_n^+\cos(\omega_n^ct) + iu\sum_{n=0}^{\infty}A_n^-\cos(\omega_n^Zt),\\
                \label{V_vx(t)}
\bar{v}_x(t) &=& u \sum_{n=0}^{\infty} B_n^+\cos(\omega_n^ct) +  u\sum_{n=0}^{\infty}B_n^-\cos(\omega_n^Zt)
                +iu\sum_{n=0}^{\infty} T_n^+\sin(\omega_n^ct) + iu\sum_{n=0}^{\infty}T_n^-\sin(\omega_n^Zt),
\end{eqnarray} \end{widetext}
where
\begin{eqnarray} \label{V_VTAD}
V_n^{\pm}&=& \mp U_{n-1,n}^{u,u} \mp U_{n,n-1}^{u,u}   - U_{n+1,n}^{l,l}    -  U_{n,n+1}^{l,l},   \nonumber \\
T_n^{\pm}&=& \pm U_{n-1,n}^{u,u} \mp U_{n,n-1}^{u,u}   + U_{n+1,n}^{l,l}    -  U_{n,n+1}^{l,l},   \nonumber \\
A_n^{\pm}&=&   - U_{n,n}^{u,l}    +  U_{n,n}^{l,u}   \pm U_{n-1,n+1}^{u,l} \mp U_{n+1,n-1}^{l,u}, \nonumber \\
B_n^{\pm}&=&     U_{n,n}^{u,l}    +  U_{n,n}^{l,u}   \pm U_{n-1,n+1}^{u,l} \pm U_{n+1,n-1}^{l,u},
\end{eqnarray}
in which
\begin{equation} \label{V_def_U}
U^{\alpha,\beta}_{m,n} = \int F_{m}^{\alpha *}(k_x) F_{n}^{\beta}(k_x) dk_x.
\end{equation}
The superscripts $\alpha, \beta$ refer to the upper and lower components, see Eqs. (\ref{V_Matrix_FuFl}).
The velocity averages must be real values. For example, if both $g^j(k_x,y)$ ($j=u,l$) in
Eq. (\ref{V_Matrix_gj}) are real then also both  $F_n^j(k_x)$ given by Eq. (\ref{V_Matrix_Fj}) are real,
and there is $U^{\alpha,\beta}_{m,n}=U^{\beta,\alpha}_{n,m}$.
As a result the last two terms in Eqs. (\ref{V_vy(t)}) and (\ref{V_vx(t)}) vanish.

The time dependent sine and cosine functions come from
the exponential terms in Eq. (\ref{V_vi(t)}).
The frequencies in Eqs. (\ref{V_vy(t)}) and (\ref{V_vx(t)})
are  $\omega_n^c=\Omega(\sqrt{n+1}-\sqrt{n})$,
$\omega_n^Z=\Omega(\sqrt{n+1}+\sqrt{n})$, where $\Omega$ is given in Eq. (\ref{H aap}).
 The frequencies $\omega_n^c$ correspond to the
intraband (cyclotron) energies while frequencies $\omega_n^Z$ correspond to the interband energies, see Fig. 1.
The interband frequencies are characteristic of the Zitterbewegung because the trembling motion
is caused by an interference of states corresponding to the positive and negative energies
\cite{BjorkenBook,GreinerBook}.
The intraband (cyclotron) energies are due to the band quantization by the magnetic field and they
do not appear in field-free situations (see \cite{Schliemann05,Zawadzki05KP,Zawadzki06,Rusin07b}).

\begin{figure}
\includegraphics[width=8.5cm,height=8.5cm]{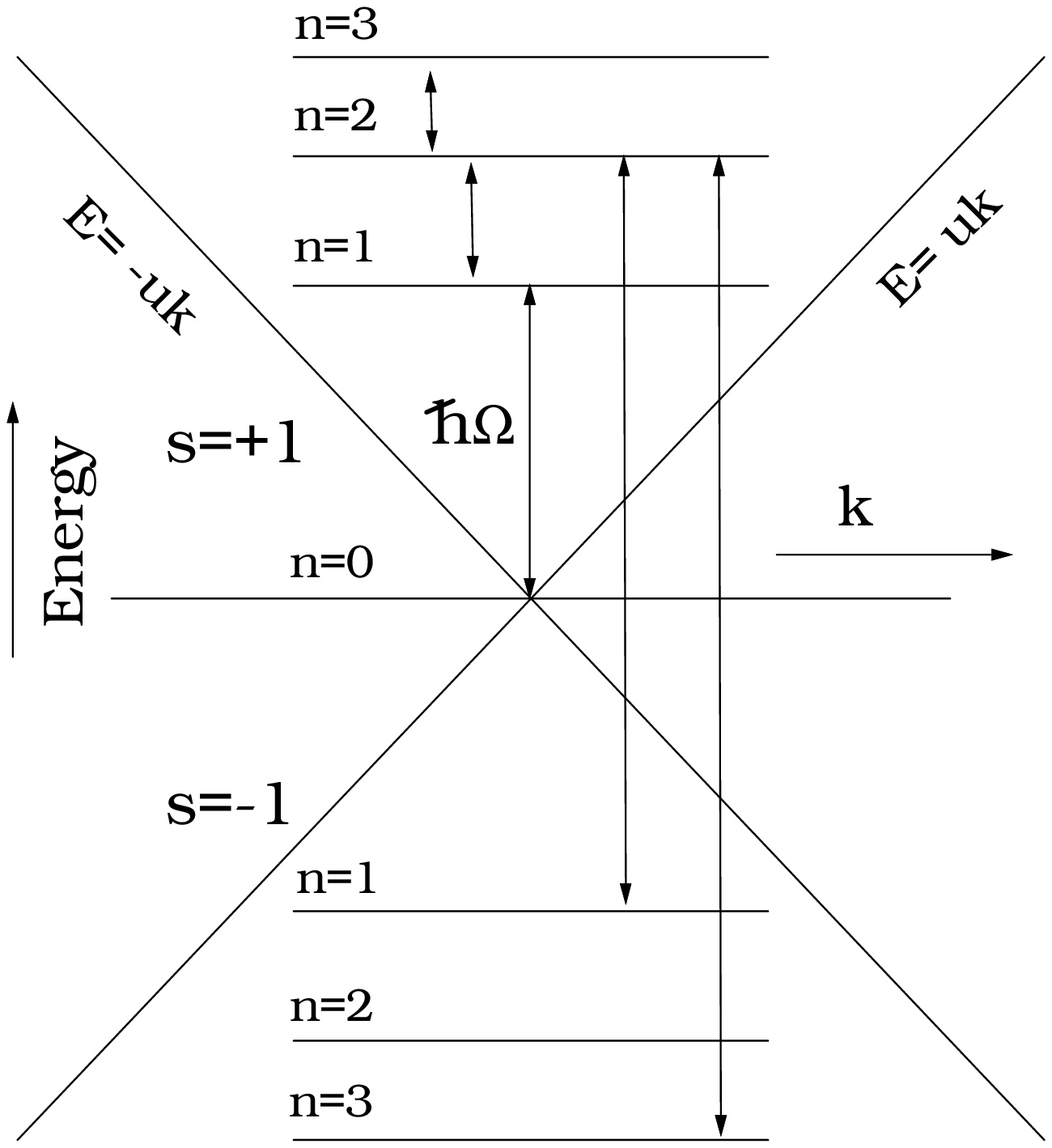}
\caption{ \label{Fig1} The energy dispersion $E(k)$ and the Landau levels for graphene in a
magnetic field (schematically). Intraband (cyclotron) and interband energies for $n'=n\pm 1$
are indicated, see text. The basic energy is $\hbar\Omega=\sqrt{2}\hbar u/L$.}
\end{figure}

\section{Gaussian wave packet}
We take the function $f(x,y)$ in the form of a Gaussian wave packet having an initial nonvanishing
momentum $p_{0x}=\hbar k_{0x}$
\begin{equation} \label{Gauss_f}
f(x,y) = \frac{1}{\sqrt{\pi d_xd_y}}e^{-\frac{x^2}{2d_x^2}-ik_{0x}x-\frac{y^2}{2d_y^2}}
         \left(\begin{array}{c} a_u \\ a_l \end{array}\right).
\end{equation}
In the above model the upper and lower components of $f$ differ only by the coefficients $a_u$ and $a_l$,
respectively. There is $a_u^2+a_l^2=1$. Then  [see Eq. (\ref{V_Matrix_gj})]
\begin{equation} \label{Gauss_g}
g(k_x,y) = \sqrt{\frac{d_x}{\pi d_y}}e^{-\frac{1}{2}d_x^2(k_x-k_{0x})^2} e^{-\frac{y^2}{2d_y^2}}
           \left(\begin{array}{c} a_u \\ a_l \end{array}\right).
\end{equation}
This gives [see Eq. (\ref{V_Matrix_Fj})]
\begin{equation} \label{Gauss_Fn_wyn}
 F^{\cal \alpha}_n(k_x) = \frac{a_\alpha A_n\sqrt{L d_x}}{\sqrt{2\pi d_y}C_n} e^{-\frac{1}{2}d_x^2(k_x-k_{0x})^2}
            e^{-\frac{1}{2}k_x^2D^2}\ {\rm H}_n(k_xc).
\end{equation}
Here $D=L^2/\sqrt{L^2+d_y^2}$, $c=L^3/\sqrt{L^4-d_y^4}$, and
\begin{equation} \label{Gauss_An}
 A_n=\frac{\sqrt{2\pi}d_y}{\sqrt{L^2+d_y^2}}\left(\frac{L^2-d_y^2}{L^2+d_y^2}\right)^{n/2}.
\end{equation}
After some manipulations, we finally find (see \cite{PrudnikovBook})
\begin{eqnarray} \label{Gauss_U_mn}
U_{m,n}^{\alpha,\beta}= \frac{a_{\alpha}a_{\beta} A_m^* A_nLQd_x \sqrt{\pi}\ e^{-W^2}}{2\pi C_mC_n d_y}
    \sum_{l=0}^{\min\{m,n\}}\!\!\!\! 2^l l! \! \left(\begin{array}{c} m \\ l\!\! \end{array}\right) \!\!\!
                                         \left(\begin{array}{c} n \\ l\!\! \end{array}\right)
   &&    \nonumber \\
  \times \left(1-(cQ)^2\right)^{(m+n-2l)/2}
    {\rm H}_{m+n-2l} \left(\frac{cQY}{\sqrt{1-(cQ)^2}}\right),  \ \ \ \ \ &&
\end{eqnarray}
where  $Q = 1/\sqrt{d_x^2+D^2}$, $W= d_xDQ$, and $Y=d_x^2k_{0x}Q$.
Thus in case of
a Gaussian wave packet we obtain the coefficients $U_{m,n}^{\alpha,\beta}$ in the
form of analytical sums.  For the special case $d_y=L$, there is simply
\begin{eqnarray} \label{Gauss_U_mn_Ldy}
U_{m,n}^{\alpha,\beta} &=& \frac{\sqrt{\pi}\ i^{m+n}\ a_{\alpha}a_{\beta}d_x}{C_mC_nL}
   \left(\frac{L}{2P} \right)^{m+n+1} \times \nonumber \\
 &&  \exp\left(-\frac{d_x^2k_{0x}^2L^2}{2P^2}\right)
     {\rm H}_{m+n}\left(\frac{-id_x^2k_{0x}}{P}\right), \ \ \
\end{eqnarray}
where $P=\sqrt{d_x^2+\frac{1}{2}L^2}$. In the above model the coefficients $U_{m,n}^{\alpha,\beta}$
are real numbers, so that $A_n^{\pm}$ and $T_n^{\pm}$ in Eqs. (\ref{V_vy(t)}) and (\ref{V_vx(t)})
vanish. A sum rule for $U_{n,n}^{\alpha,\beta}$ is given in Appendix A.

\section{Results and Discussion}

\begin{figure}
\includegraphics[width=8.5cm,height=11.33cm]{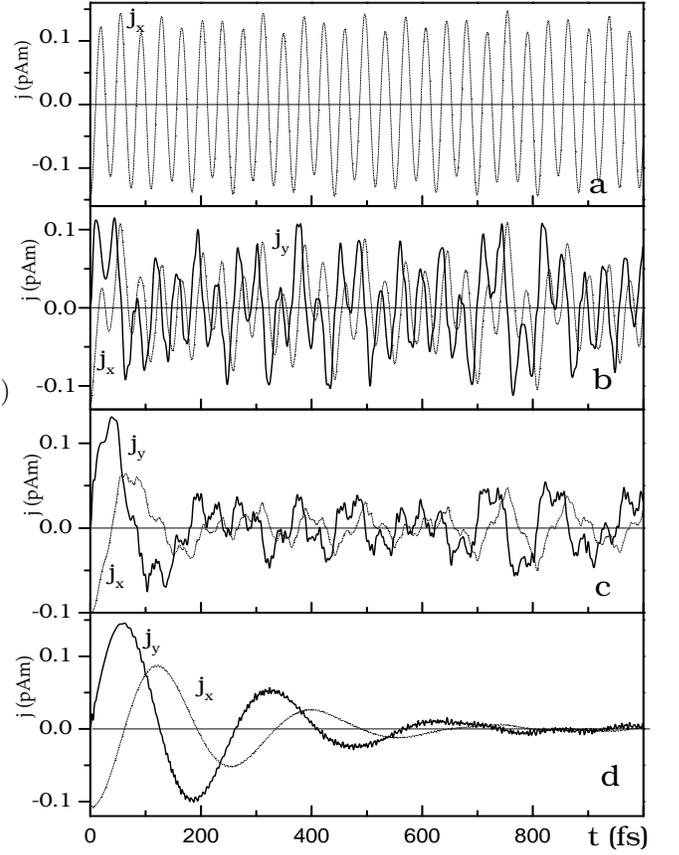}
\caption{ \label{Fig2} Contribution of the $K_1$ point of the Brillouin zone
to the electric current in graphene
at B=10T versus time, calculated for a Gaussian wave packet of the width $d_x=d_y=L=81.13$\AA\ and
various values of $k_{0x}$:
a) $k_{0x}=0$, b) $k_{0x}=0.02$\AA$^{-1}$, c) $k_{0x}=0.035$\AA$^{-1}$, d) $k_{0x}=0.06$\AA$^{-1}$.
Thick lines -- $j_y(t)$, thin lines -- $j_x(t)$.}
\end{figure}

\begin{figure}
\includegraphics[width=8.5cm,height=8.5cm]{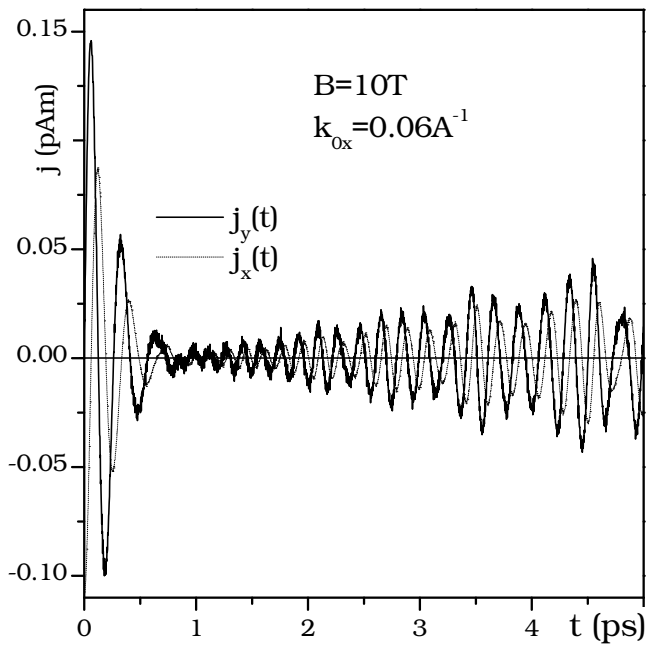}
\caption{ \label{Fig3} The same as Fig. 2d, but for larger time scale.
Results for the first picosecond coincide with those shown in Fig. 2d.}
\end{figure}

\begin{figure}
\includegraphics[width=8.5cm,height=8.5cm]{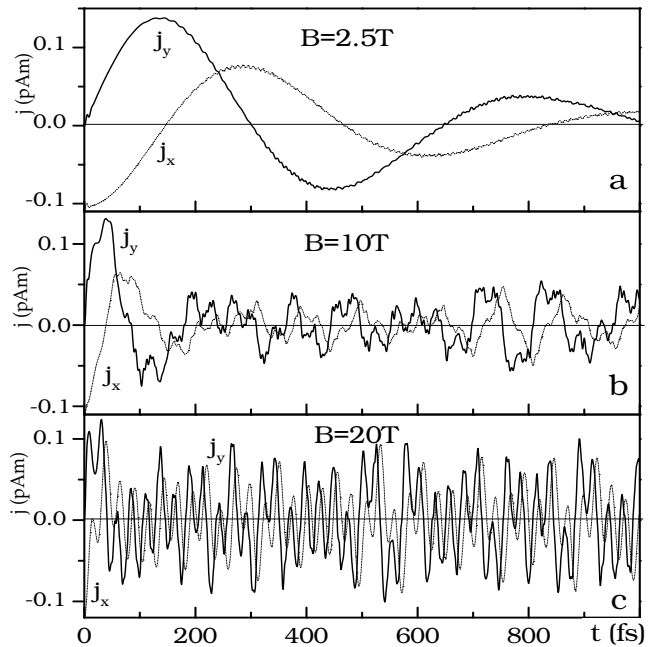}
\caption{ \label{Fig4} The same as in Fig. 2, calculated for fixed values of $k_{0x}=0.035$\AA$^{-1}$ and
widths $d_x=d_y=81.13$\AA\, but at different magnetic fields.
Results 4b are the same as those shown in Fig. 2d.}
\end{figure}

\begin{figure}
\includegraphics[width=8.5cm,height=8.5cm]{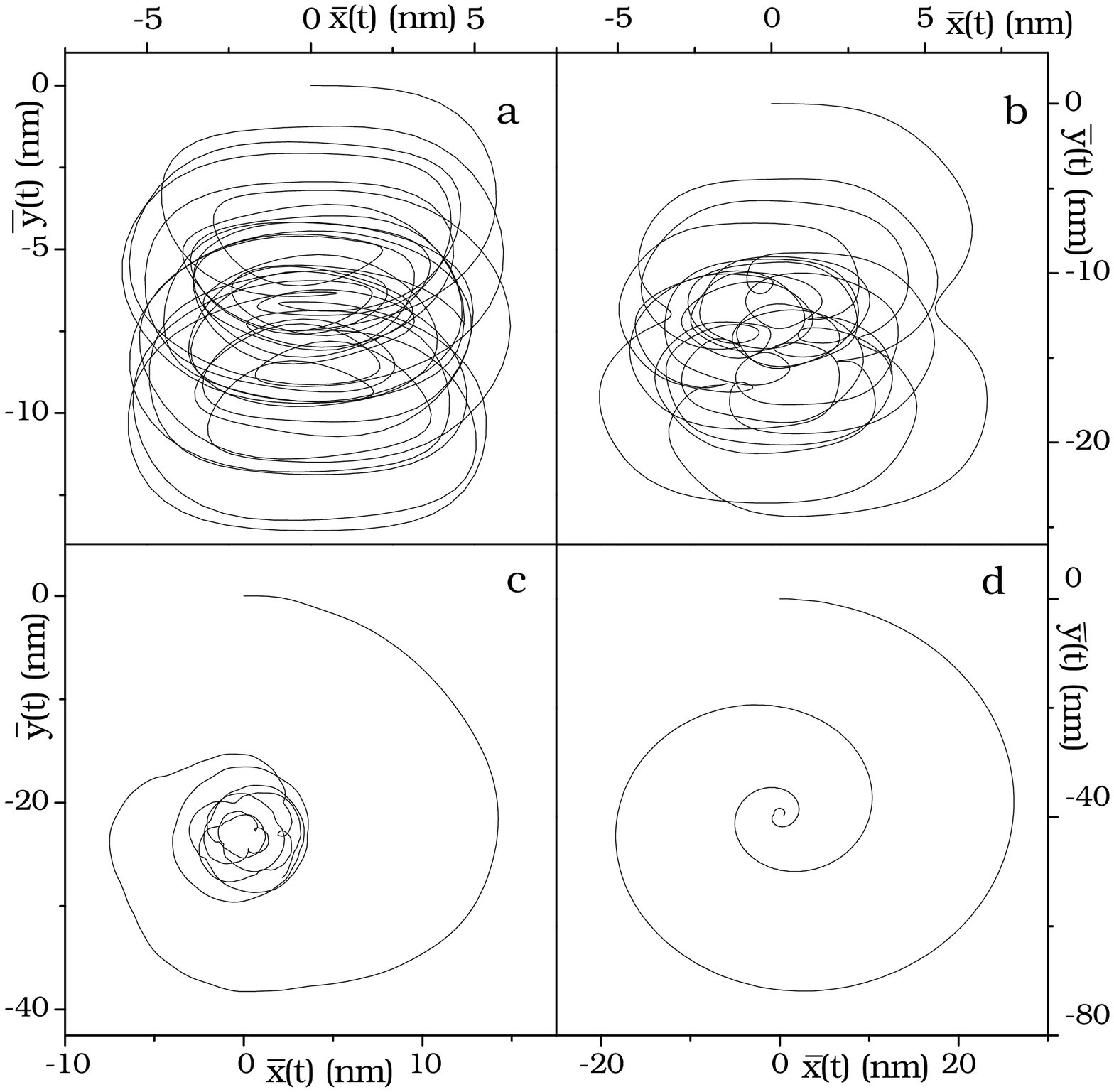}
\caption{ \label{Fig5} Zitterbewegung trajectories of electron at the $K_1$ point of the Brillouin
zone during the first picosecond for various values of $k_{0x}$:
a) $k_{0x}=0.01$\AA$^{-1}$, b) $k_{0x}=0.02$\AA$^{-1}$ c), $k_{0x}=0.035$\AA$^{-1}$ d),
$k_{0x}=0.06$\AA$^{-1}$.}
\end{figure}

In Fig. 2 we show calculated current components $j_i=-e\bar{v}(t)$ as
functions of time for different values of the initial wave vector $k_{0x}$.
The calculations were carried out for the $K_1$ point of BZ taking $a_u=a_l=1/\sqrt{2}$ and
a constant magnetic field of 10T. We assumed a circular wave packet $d_x=d_y=L=83.13$\AA,
the coefficients $U_{m,n}^{\alpha,\beta}$ were calculated using Eq. (\ref{Gauss_U_mn_Ldy}).
Figure 2a shows the results for $k_{0x}=0$.
It can been seen that, if there is no initial momentum, only $j_x(t)$
is nonzero. The main frequency of oscillations is $\omega_0=\Omega$, which can be
interpreted either as $\omega_0^c=\Omega(\sqrt{n+1}-\sqrt{n})$ or
$\omega_0^Z=\Omega(\sqrt{n+1}-\sqrt{n})$
for $n=0$. Frequency $\omega_0^c$ belongs to the intraband (cyclotron) set, while $\omega_0^Z$
belongs to the interband set (see Fig. 1). Somewhat unequal amplitude of oscillations means
that other frequencies also come into play, so that we already deal with the trembling motion.
For $k_{0x}=0$, the asymmetry between $\bar{v}_x(t)$
and $\bar{v}_y(t)$ comes from the above mentioned asymmetry of the initial Hamiltonian
with respect to $\hat{p}_x$ and  $\hat{p}_y$, see Eq. (\ref{H_pi0}).
In Figs. 2b, 2c and 2d we show calculated contributions to the current
for growing values of $k_{0x}$. For nonzero values of $k_{0x}$, both $j_x$ and $j_y$ appear.
It is seen  that the frequency and the shape of ZB
oscillations change with growing $k_{0x}$ values.
For growing $k_{0x}$ different $U_{m,n}^{\alpha,\beta}$ become large
and in consequence different frequencies $\omega_n^c$ and $\omega_n^Z$ dominate in sums
(\ref{V_vy(t)}) and (\ref{V_vx(t)}). The striking feature seen in Figs. 2a, 2b and 2c is, that the ZB is
manifested by several frequencies simultaneously. This is a consequence of the fact that, as follows
from Eq. (\ref{H_Ens}), in graphene the energy distances between the Landau levels diminish with $n$,
which results in different values of frequencies  $\omega_n^c$ and $\omega_n^Z$ for different $n$.
Thus, it is the presence of an external quantizing magnetic field that introduces various frequencies
into ZB.

For sufficiently large values of $k_{0x}$ only {\it one} frequency prevails, as shown in Fig. 2d.
This is related to the fact that, as seen in the inset of Fig. 8 in Appendix B,
the coefficients $U_{m,n}^{\alpha,\beta}$ in this regime have a pronounced maximum around a
specific value $n_{max}$. The dominant frequency is $\omega_{max}=\omega(\sqrt{n_{max}+1}-\sqrt{n_{max}})$,
which is simply the cyclotron frequency for $n_{max}$, see Fig. 1. Thus, it might appear that
the current shown in Fig. 2d corresponds to the simple classical cyclotron motion and
the trembling motion is manifested only by the damping
in time (see Appendix D). This is however not the case.

In Fig. 3 we show the calculated current for $k_{0x}=0.06$\AA$^{-1}$ (the same as in Fig. 2d), but
in a much larger time scale. It turns out that, after the ZB oscillations seemingly die out, they
actually reappear.

Thus, {\it for all $k_{0x}$ values (including $k_{0x}=0$), the ZB oscillations have a permanent character},
that is they do not disappear in time. This feature is due to the discrete character of the electron spectrum
caused by a magnetic field. The above property is in sharp contrast to the no-field cases
considered until present, in which the spectrum is not quantized and the ZB of a wave packet
has a transient character, see \cite{Rusin07b,Zawadzki08}.
In mathematical terms, due to the discrete character of the spectrum,
averages of operator quantities taken over a wave packet are sums and not integrals, see
Eqs. (\ref{V_vy(t)}) and (\ref{V_vx(t)}). The sums do not obey the Riemann-Lebesgues theorem for
integrals which guaranteed the damping of ZB in time for a continuous spectrum (see Ref. \cite{Lock79}).
We consider the demonstration of a permanent character of ZB oscillations for
a discrete spectrum to be the main result of our present work.

In Fig. 4 we show the ZB oscillations of the current for a constant wave vector $k_{0x}=0.035$\AA$^{-1}$ at
different magnetic fields. It can be seen that the intensity of a magnetic field has a dramatic
effect on ZB: not only its frequency is changed but also the character of oscillations. Lower
magnetic fields are equivalent to higher $k_{0x}$ values, since both lead to higher Landau levels
involved (see Figs. 2d and 4a). Inversely, higher magnetic fields and lower $k_{0x}$ values lead to
lower Landau levels involved (see Figs. 2b and 4c).
At very small magnetic fields there exist three regimes of ZB oscillations: the 'initial' oscillations
dying out during a few femtoseconds (as shown in Fig. 2 of Ref. \cite{Rusin07b}), the second range
of oscillations dying out during several picoseconds, as shown in Fig. 2d, and the third range of
permanent, somewhat irregular oscillations shown for $t\ge 1$ps in Fig. 3.

Finally, we calculate the displacements $\bar{x}(t)$ and $\bar{y}(t)$ of the wave packet. To this end we
integrate expressions (\ref{V_vy(t)}) and (\ref{V_vx(t)}) with respect to time using the initial conditions
$x_0=\bar{x}(0)$ and $y_0=\bar{y}(0)$. The results are plotted in Fig. 5 in the form of  $x-y$ trajectories
for different initial wave vectors $k_{0x}$. The direction of movement is clockwise.
The trajectories span early times (1ps) after the creation of a wave packet. As mentioned above,
the ZB oscillations do not die out in time which is reflected by infinite trajectories. In Fig. 5 the
trajectories are shown around the point $x_0=y_0=0$, whereas in reality the $y$ component of
the center is almost a linear function of $k_{0x}: y_0\approx k_{0x}L^2$.

All in all, the presence of a quantizing magnetic field has the following important effects on the
trembling motion. (1) For $B\neq 0$ the ZB oscillation are permanent, while for $B=0$ they are transient.
(2) For $B\neq 0$ many ZB frequencies appear, whereas for $B=0$ only one ZB frequency exists.
(3) For $B\neq 0$ both interband and intraband (cyclotron) frequencies appear in ZB; for $B=0$ there
are no intraband frequencies. (4) Magnetic field intensity changes not only the ZB frequencies
but the entire character of ZB spectrum.

The results shown in Fig. 2 were obtained using the simplifying assumption about packet's
width: $d_x=d_y=L=(\hbar/eB)^{1/2}$. This allowed us to use formula (\ref{Gauss_U_mn_Ldy})
for the calculation of $U_{m,n}^{\alpha,\beta}$.
However, in Fig. 4 we show the results obtained for constant $d_x=d_y$
and variable $B$, for which we had to use general formula (\ref{Gauss_U_mn}).
In all the calculations involving magnetic field, precise numerical
values of the Hermite polynomials are required, see Eqs. (\ref{Gauss_U_mn}) and (\ref{Gauss_U_mn_Ldy}).
For the results shown above we used the values of the first 400 Hermite polynomials,
and checked their high precision using
sum rule (\ref{Gauss_A1}). We also considered the case of an initial electron momentum
directed not along the $x$ direction (as shown above), but also along the $y$ direction. In this case
the $U_{m,n}^{\alpha,\beta}$ coefficients are imaginary, so that only the last two terms of
Eqs. (\ref{V_vy(t)}) and (\ref{V_vx(t)}) come into play.
The final results are similar but not identical to those quoted
above. The reason is that, as already mentioned, the initial Hamiltonian (\ref{H_pi0}) is not symmetric
in $\hat{p}_x$ and $\hat{p}_y$ momenta. When using Gaussian  wave packet (\ref{Gauss_f}), we assumed
equal upper and lower components $a_u=a_l=1/\sqrt{2}$. This is in contrast to previous papers which usually
took $a_u=1$, $a_l=0$ \cite{Schliemann05,Rusin07a,Rusin07b,Lock79,Huang52}. This choice is somewhat arbitrary,
it is determined by an experimental wave packet usually prepared by optical methods. One should keep in mind that
the relative final amplitudes of $\bar{v}_x(t)$ and $\bar{v}_y(t)$ (and the resulting currents) depend on
this choice via $U_{m,n}^{\alpha,\beta}$ coefficients, see Eqs. (\ref{Gauss_U_mn}) and (\ref{Gauss_U_mn_Ldy}).
If one chooses $a_u=1$, $a_l=0$, the resulting motion coming from the $K_1$ point of BZ is only
along the $y$ direction.

It is of interest that phenomena analogous to those described above for electrons, occur also for
photons. In particular, Hamiltonian (\ref{H aap}) is similar to that describing an interaction of atoms
with electromagnetic radiation according to the so called Jaynes-Cummings model \cite{Jaynes63,GerryBook}.
In particular, the collapse and revival of electron's ZB oscillations,
as illustrated in Figs. \ref{Fig2}b and \ref{Fig3},
is predicted by the Jaynes-Cummings model for the number of emitted photons ant is was observed in
one-electron masers \cite{Rempe87}.

\section{Dipole radiation due to ZB}
\begin{figure}
\includegraphics[width=8.5cm,height=8.5cm]{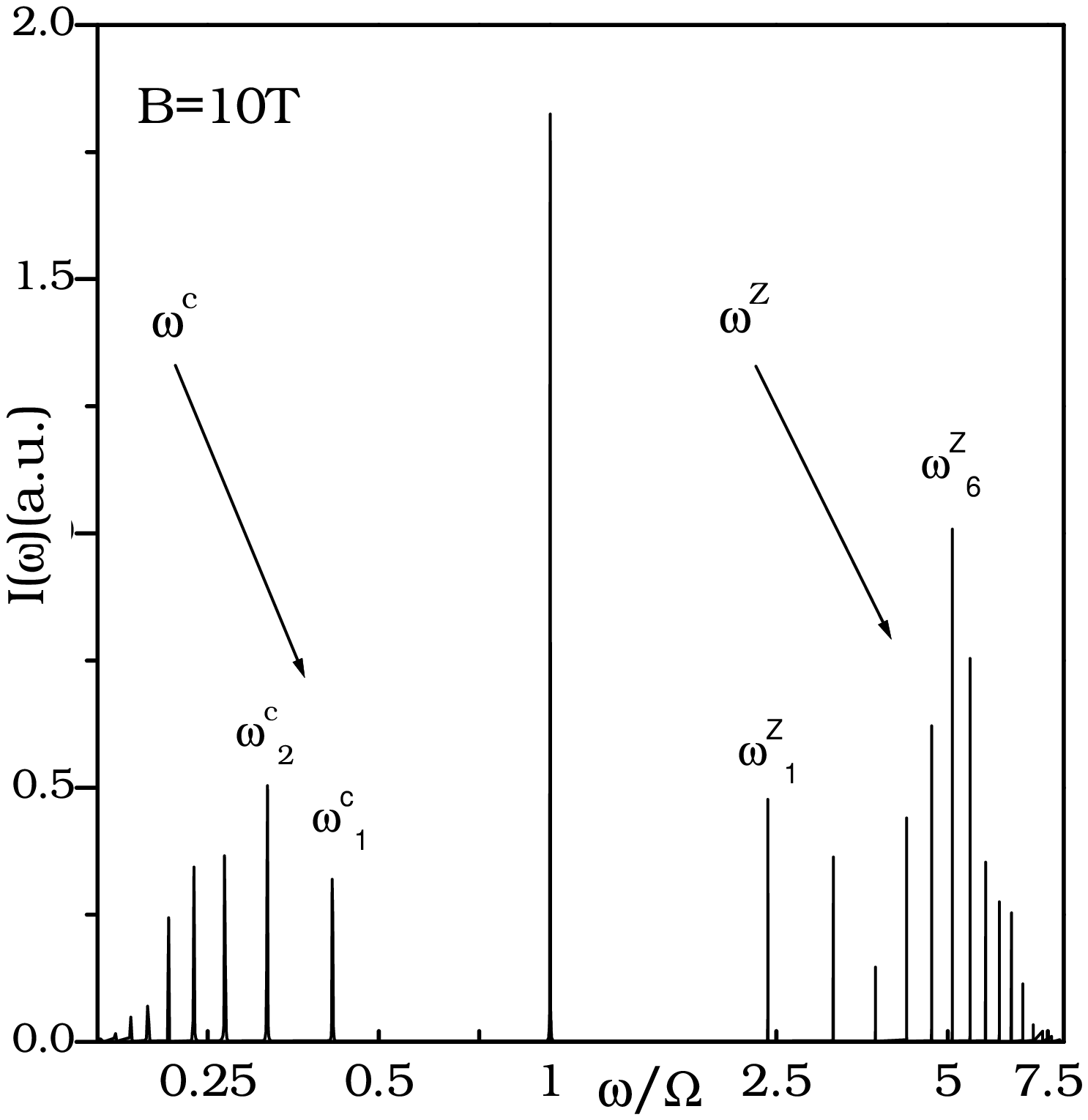}
\caption{ \label{Fig6} Intensity spectrum versus frequency during the first 20 ps of motion of an
electron described by a Gaussian wave packet having
$k_{0x}=0.035$\AA$^{-1}$ at $B=10$ T. The frequencies $\omega_n^c$ and $\omega_n^Z$ are defined in
Eqs. (\ref{V_vy(t)}) and (\ref{V_vx(t)}).}
\end{figure}

Experimental possibilities of observing the trembling motion were considered previously
\cite{Schliemann05,Zawadzki06,Rusin07b} and we do not consider this problem again. The results shown
in Figs. 2, 3 and 4 describe the electric current, which is an observable quantity. One could also
try to see directly the displacement of charge, cf. \cite{Topinka00,LeRoy03}.
On the other hand, we argue below that the ZB should be accompanied by electromagnetic dipole
radiation emitted by the trembling electrons.
The oscillations $\bar{r}(t)$, as shown in Fig. 5, are related to the dipole moment $-e\bar{r}(t)$,
which couples to the electromagnetic radiation. We shall treat the radiation classically \cite{BohmBook}, i.e. we
take the radiated transverse electric field to be \cite{JacksonBook}
\begin{equation} \label{Dip_E}
 {\cal \bm E}_{\perp}(\bm r,t) = \frac{e\bar{\ddot{\bm r}}(t)}{4\pi\epsilon_0c^2}
                          \frac{\sin(\theta)}{R},
\end{equation}
where $\epsilon_0$ is the vacuum permittivity, $\theta$ is an angle between the direction of motion
of a wave packet and the position of the observer $\bm R$.
Integrating ${\cal \bm E}_{\perp}^2$ over the angle $\theta$ one obtains the total
radiated power given by the Larmor formula
\begin{equation} \label{Dip_P}
P = \frac{e^2\bar{a}^2(t)^2}{6\pi\epsilon_0c^3},
\end{equation}
where $\bar{a}$ is the acceleration averaged over the packet.
To find $P$ we calculate the acceleration from Eqs. (\ref{V_vy(t)}) and (\ref{V_vx(t)})
by taking the time derivatives. The spectrum of the emitted radiation is obtained by the Fourier
transform of the electric field. We have
\begin{equation}
\vec{\cal E}(t) = \frac{1}{2}a_0 + \sum_{m=0}^{\infty} \vec{a}_m \cos\left(\frac{m\pi t}{T}\right) + \vec{b}_m\sin\left(\frac{m\pi t}{T}\right),
\end{equation}
where
\begin{eqnarray}
\vec{a}_m = \lim_{T\rightarrow \infty} \int_{-T}^T \cos\left(\frac{m\pi t}{T}\right)\vec{\cal E}_{\perp}(\bm r,t) dt, \nonumber \\
\vec{b}_m = \lim_{T\rightarrow \infty} \int_{-T}^T \sin\left(\frac{m\pi t}{T}\right)\vec{\cal E}_{\perp}(\bm r,t) dt,
\end{eqnarray}
and $a_0=0$. For the numerical calculations we take a large period $T=20$ ps.
The intensity spectrum of oscillations is
\begin{equation}
I(\omega_m) \propto \sum_m (\vec{a}_m^2 +  \vec{b}_m^2).
\end{equation}
The plot of $I(\omega_m)$ is given in Fig. \ref{Fig6}.
The strongest frequency peak corresponds to oscillations with the basic frequency $\omega=\Omega$.
The peaks on the high frequency side correspond to the interband excitations and are characteristic of ZB.
The peaks on the lower frequency side correspond to the intraband (cyclotron) excitations.
At  higher magnetic fields we may expect smaller number of distinct frequencies,
while for lower fields the classical radiation will evolve toward a quasi-continuous spectrum.
In absence of Zitterbewegung the emission spectrum would contain only the intraband (cyclotron)
frequencies (see Appendix D). Thus the interband frequencies $\omega_n^Z$ shown in Fig. \ref{Fig6}
are a direct signature of the trembling motion. It can be seen that the  $\omega_n^Z$ peaks are not
drastically weaker than the central peak at $\omega=\Omega$ which means that there exists a reasonable
chance to observe them.

\begin{figure}
\includegraphics[width=8.5cm,height=8.5cm]{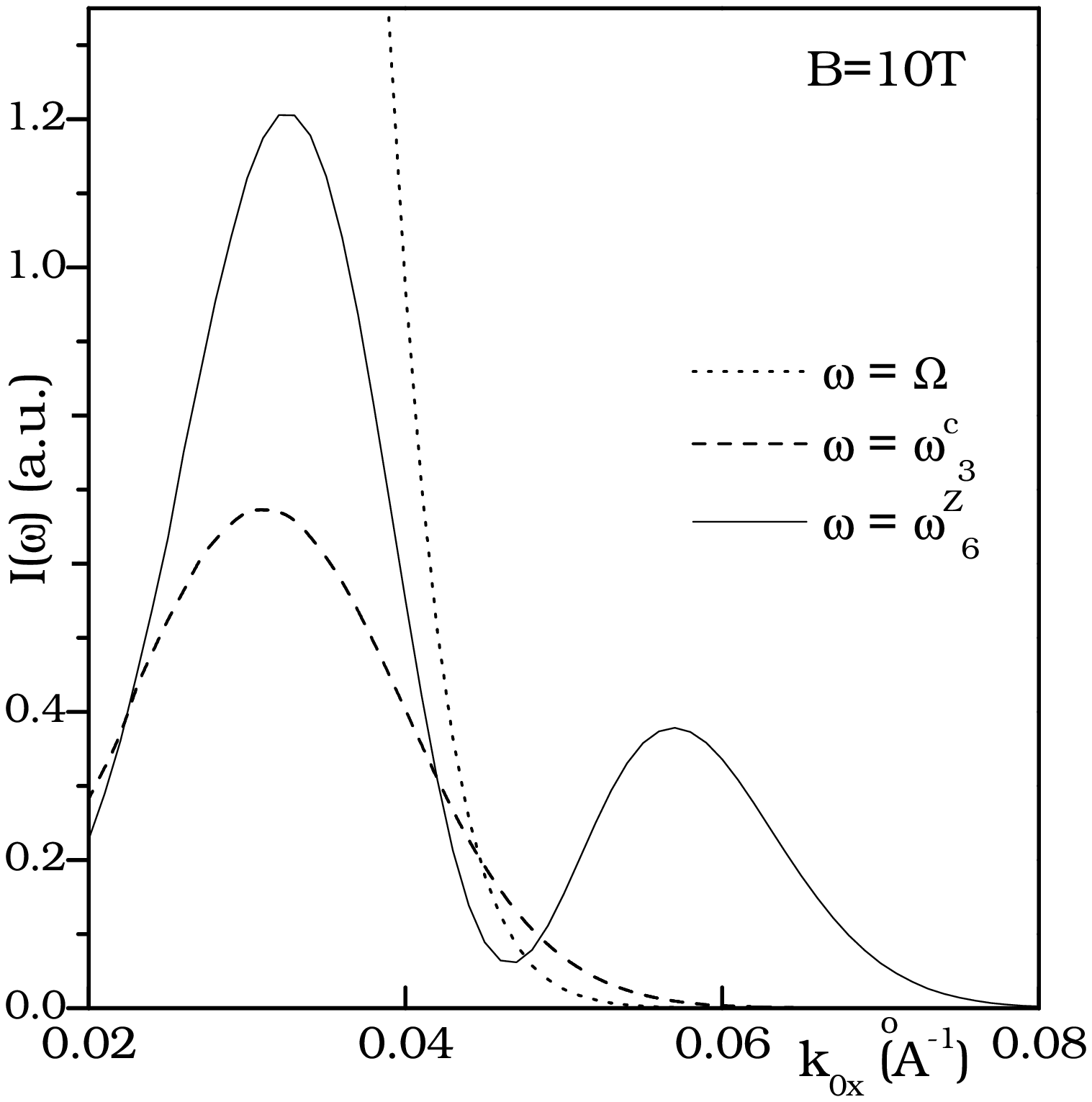}
\caption{ \label{Fig7} Intensities of emission lines corresponding to $\omega^c_3$ (dashed line),
$\omega_0=\Omega$ (dotted line) and to $\omega^Z_6$ (solid line) versus the wave vector $k_{0x}$
of Gaussian wave packet at $B=10$ T.}
\end{figure}

In Fig. \ref{Fig7} we plot the dependence of the emitted power intensity on the initial wave vector
$k_{0x}$ for three lines (calculated for $B=10$ T): the basic cyclotron line at $\omega=\Omega$,
the intraband frequency $\omega_3^c$, and the interband frequency $\omega_6^Z$. It is seen
that the intensity of various emission lines depends differently on $k_{0x}$.
At small $k_{0x}$ values the basic line $\omega=\Omega$ dominates,
but at $k_{0x}\simeq 0.04$\AA$^{-1}$ the intensities of
various lines become comparable. The characteristic two maxima of $\omega_6^Z$ occur also for the
other interband frequencies. We believe that the $k_{0x}$ dependence of the line intensities,
as shown in Fig. \ref{Fig7}, can serve as a signature of ZB.

The properties shown in Figs. \ref{Fig2} - \ref{Fig7} have been calculated for the $K_1$ point
of the Brillouin zone in graphene. The main features applying to the $K_2$ point of BZ are
described in Appendix E. The above calculations are somewhat idealized since they do not
take into account the position of the Fermi energy in a given sample.
Thus they correspond approximately to a
situation with the Fermi energy relatively low in the valence band. Clearly, the frequencies
corresponding to transitions with the final states below the Fermi energy are not possible.

In general terms, the excitation of the system we propose is due to the nonzero momentum
$\hbar k_{0x}$ given to the electron. It can be provided by accelerating the
electron in the band by or by exciting the electron with a nonzero momentum by light
from the valence band to the conduction band. The electron can emit light because
the Gaussian wave packet is not an eigenstate of the system described by
Hamiltonian (\ref{H_pi0}). The energy of the emitted light is provided by the initial
kinetic energy related to the momentum $\hbar k_{0x}$. Once this energy
is completely used the emission will cease. If the electron is described by a
non-gaussian wave packet, all our results are quantitatively valid, only the
intensity spectrum will differ from that shown in Fig. \ref{Fig6}.
We emphasize that the sustained character of ZB oscillations caused by
a discrete energy spectrum makes graphene in a magnetic field probably
the most favorable system for an experimental observation of the
trembling motion considered until present.

\section{Summary}
We described the Zitterbewegung of electrons in solids in the presence of a magnetic field
assuming that the electrons are represented by Gaussian wave packets. The system under consideration is
monolayer graphene. It is shown that the presence of a quantizing magnetic field has a profound influence
on the Zitterbewegung. In particular, the discrete energy spectrum in a magnetic field causes the Zitterbewegung
to be sustained in time while for $B=0$ the ZB has a transient character. In addition, at $B\neq 0$ many
ZB frequencies appear whereas at $B=0$ one deals with only one ZB frequency. For a given value of initial
electron momentum, the magnetic field intensity affects not only ZB frequencies but the entire shape of the ZB
spectrum. We consider and describe an electromagnetic radiation emitted by the trembling electrons.
Graphene in a magnetic field seems to be a very favorable system for an experimental
observation of Zitterbewegung.

\appendix
\section{}
We consider a sum rule for the coefficients $U_{m,n}^{\alpha,\beta}$ of Eq. (\ref{V_def_U}).
Let us calculate $1=\langle f|f\rangle$,
\begin{eqnarray} \label{Gauss_A1}
1&=& \sum_{\rm n} |\langle f|{\rm n}\rangle|^2 =
    \sum_{n,s} \int dk_x |-sF_{n-1}^u(k_x)+F_{n}^l(k_x)|^2 \nonumber \\
 &=&2\sum_{n=0}(U_{n,n}^{u,u}+ U_{n,n}^{l,l}).
\end{eqnarray}
Factor $2$ appears due to the summation over $s$.  It is to be reminded that  $a_u^2+ a_l^2=1$.
The above sum rule can be used for a verification of the numerical values of $U_{n,n}^{\alpha,\alpha}$.

\section{}

\begin{figure}
\includegraphics[width=8.5cm,height=8.5cm]{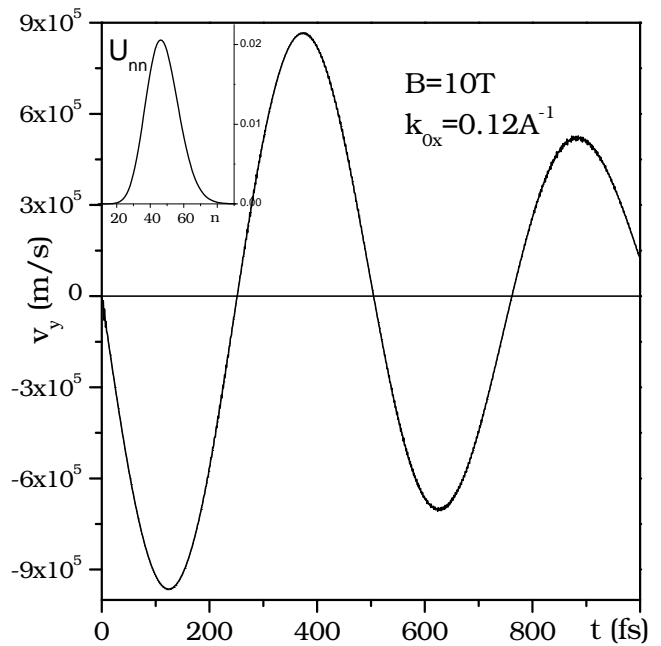}
\caption{ \label{Fig8} Contribution of the $K_1$ point to the electron velocity $\bar{v}_y(t)$
versus time, calculated for the indicated parameters: (a) using full formula (\ref{V_vy(t)}) (see
Fig. 2d), (b) using the first integral of the Poisson formula (\ref{V_vy(t)Poisson}).
The two curves practically coincide. Inset shows coefficients $U_{n,n\pm 1}=U_{n\pm 1,n} \approx U(x)$
for $k_{0x}=0.12$\AA$^{-1}$. The frequency of oscillations corresponds to
$\omega_{max}$ for $x_{max}=46$, see text.}
\end{figure}

We calculate $\bar{v}_y(t)$  from Eq. (\ref{V_vy(t)}) for a situation when $U_{n,n}^{\alpha,\beta}$ have a
maximum for a large value of $n$. In our model there is $U_{n,n}^{\alpha,\beta}=a_{\alpha}a_{\beta}U_{n,n}$.
We use the Poisson summation formula for an estimation of the velocity average disregarding
$a_{\alpha}$ and $a_{\beta}$ coefficients. Upon replacing $U_{m,n}$ by a continuous variable $U(x)$
and approximating $U_{n,n\pm 1}=U_{n\pm 1,n} \approx U(x)$,
the term with $V_n^-$ in Eq. (\ref{V_vy(t)}) vanishes. Then
\begin{eqnarray} \label{V_vy(t)Poisson}
\bar{v}_y(t)\approx 4u\int_0^{\infty} \sin(\omega_x t)U(x)dx  \nonumber \\
            +4u\sum_{l=1}^{\infty} \int\sin(\omega_x t)U(x)\cos(2\pi x l)dx,
\end{eqnarray}
where $\omega_x = \Omega(\sqrt{x+1}-\sqrt{x})$. For sufficiently small times
we may disregard the second term and $\bar{v}_y(t)$ is given by the first integral in
Eq. (\ref{V_vy(t)Poisson}).
In Fig. \ref{Fig8} we show the results of the integration compared
with the exact calculations of Eq. (\ref{V_vy(t)}) for $k_{0x}=0.12$\AA$^{-1}$.
The two curves practically coincide, apart from the small contributions of higher ZB frequencies
present in exact formula (\ref{V_vy(t)}). The effective frequency of the motion
is given by $\omega_{max}=\Omega(\sqrt{x_{max}+1}-\sqrt{x_{max}})$, where $x_{max}= 46$
corresponds to the maximum of $U(x)$, see inset.  For larger
times the second term in Eq. (\ref{V_vy(t)Poisson}) is not negligible and full
formula (\ref{V_vy(t)Poisson}) is equivalent to Eq. (\ref{V_vy(t)}).

\section{}
Here we consider briefly the gauge aspects. According to the general theory \cite{Kobe78},
if one introduces a new gauge by means of an arbitrary function $\Lambda(\bm r)$,
the new vector potential is $\bm A'=\bm A + {\bm \nabla} \Lambda$, and
the new scalar potential is $A_0'=A_0 + \partial \Lambda/\partial t$.
Then the wave function changes as
$\Psi'=e^{(ie/\hbar)\Lambda}\Psi$, and the gauge invariance for an operator
$\hat{O}=\hat{O}(\bm A, A_0)$ means
\begin{equation} \label{AGaugDef}
 \langle \Psi|\hat{O}(\bm A, A_0)|\Psi\rangle =   \langle \Psi'|\hat{O}(\bm A', A_0')|\Psi'\rangle.
\end{equation}
This leads to \begin{eqnarray} \label{AGaugDef1}
\hat{O}(\bm A', A_0') = e^{(ie/\hbar)\Lambda}\hat{O}(\bm A, A_0)e^{-(ie/\hbar)\Lambda} =
\nonumber \\
\hat{O}(\bm A, A_0) + \left[e^{(ie/\hbar)\Lambda},\hat{O}(\bm A, A_0)\right]e^{-(ie/\hbar)\Lambda}.
\end{eqnarray}
If, instead of the gauge $\bm A=(-By,0,0)$, we take $\bm A'= (0,Bx,0)$, which gives the
same magnetic field, we have
$\Lambda = xy/L^2$, so that $\Psi'(x,y) = e^{ixy/L^2}\Psi(x,y)$. Using prescription (\ref{AGaugDef1})
and calculating
\begin{eqnarray}
\left[ e^{ixy/L^2},\hat{p}_x \right] &=& -e B y e^{ixy/L^2},  \nonumber  \\
\left[ e^{ixy/L^2},\hat{p}_y \right] &=& -e B x e^{ixy/L^2},
\end{eqnarray}
one shows that relation (\ref{AGaugDef}) is satisfied also for Hamiltonian
(\ref{H_pi0}) in the new gauge.

\section{}
We consider here the motion of a wave packet in the presence of
a magnetic field according to the Schrodinger equation. For  2D Hamiltonian there
is $|{\rm n}\rangle=|nk_x\rangle$
$=e^{ik_xx}{\rm H}_{\rm n}(\xi)e^{-\xi^2/2}/(\sqrt{2\pi L}C_n)$, and $E_n=\hbar\omega_c(n+1/2)$, where
$\omega_c=eB/m$. The velocity average $\bar{v}_x(t)$ is
\begin{equation} \label{ANON_vx_def}
\bar{v}_x(t)=\frac{1}{m}\!\!\!\sum_{n,n',k_x,k_x'}\!\!\!
             \langle f|n'k_x'\rangle\langle nk_x|f\rangle e^{i\omega_c(n'-n)t}
              \langle n'k_x'|\hat{\pi}_x| nk_x\rangle,
\end{equation}
and similarly for $\bar{v}_y(t)$. Since
\begin{equation} \label{ANON_vx_1}
\langle n'k_x'|\hat{\pi}_x| nk_x\rangle =
  -\frac{\hbar \delta_{k_x,k_x'}}{\sqrt{2}L}\left(\sqrt{n}\delta_{n',n-1}+\sqrt{n+1}\delta_{n',n+1}\right),
\end{equation}
we have
\begin{eqnarray} \label{ANON_vx_2}
\bar{v}_x(t)&=&
     -\frac{\hbar}{\sqrt{2}Lm}\int dk_x\sum_{n=1}^{\infty} \langle f|n-1\rangle\langle n|f\rangle \sqrt{n}   e^{-i\omega_c t}+ \nonumber \\
  &-& \frac{\hbar}{\sqrt{2}Lm}\int dk_x\sum_{n=0}^{\infty} \langle f|n+1\rangle\langle n|f\rangle \sqrt{n+1} e^{ i\omega_c t}. \ \ \ \ \ \ \
\end{eqnarray}
There is $\sqrt{n}|n-1\rangle = \ha|n\rangle$ and $\sqrt{n+1}|n+1\rangle = \hap|n\rangle$, and we calculate
\begin{equation} \label{ANON_vx_3}
\bar{v}_x(t)= \frac{\hbar}{\sqrt{2}Lm} \int dk_x \left(\langle f\ha |f\rangle  e^{-i\omega_c t} +
 \langle f\hap|f\rangle  e^{ i\omega_c t} \right),
\end{equation}
and similarly for $\bar{v}_y(t)$.
For a one-component wave packet of Eq. (\ref{Gauss_f}) the integrals indicated in Eq. (\ref{ANON_vx_3})
can be done analytically. We finally obtain
\begin{eqnarray} \label{ANON_fin}
\bar{v}_x(t) &=&  \frac{\hbar k_{0x}}{2m} \left(-e^{-i\omega_c t} - e^{i\omega_c t} \right) = -\frac{\hbar k_{0x}}{m}\cos(\omega_c t), \nonumber \\
\bar{v}_y(t) &=& i\frac{\hbar k_{0x}}{2m} \left( e^{-i\omega_c t} - e^{i\omega_c t} \right) = -\frac{\hbar k_{0x}}{m}\sin(\omega_c t). \ \
\end{eqnarray}
Thus an electron represented by a Gaussian wave packet having the initial momentum $\hbar k_{0x}$
moves on a circular orbit with the cyclotron frequency $\omega_c$ {\it without attenuation}.
A similar result is known for a one-dimensional wave packet moving in a parabolic potential.
On the other hand, the motion illustrated in Fig. 2d is damped during the first picosecond
which is an another manifestation of Zitterbewegung.

\section{}
Here we consider contributions related to ZB of electrons at the inequivalent point $K_2$ of the
Brillouin zone. The form of Hamiltonian at the $K_2$ point is somewhat controversial, various
authors give different expressions. According to Refs. \cite{Gusynin07} and \cite{Bena07}
the Hamiltonian $\hH'$ is
\begin{equation}
\hH' = u\left(\begin{array}{cc}
     0 & -\hat{\pi}_x-i\hat{\pi}_y \\  -\hat{\pi}_x+i\hat{\pi}_y & 0 \\     \end{array}\right),
\end{equation}
i.e. $\hH' = -\hH^T$. The eigenvectors of $\hH'$ are
\begin{equation}
|nk_xs\rangle' = \frac{e^{ik_xx}}{\sqrt{4\pi}}
               \left(\begin{array}{c} |n\rangle \\s|n-1\rangle    \end{array}\right).
\end{equation}
i.e. they differ from those given by Eq. (\ref{H_nskx}).
The quantum velocity $\partial \hH' / \partial \hat{p}_x = -u\sigma_x$ and
$\partial \hH' / \partial \hat{p}_y= +u\sigma_y$. Thus, the $x$ component of the velocity changes
sign, while the $y$ component remains unchanged. Repeating the calculations  we obtain again
Eqs. (\ref{V_vy(t)}) and (\ref{V_vx(t)}), in which the coefficients are
\begin{eqnarray} \label{AV_VTAD}
\tilde{V}_n^{\pm}&=& - U_{n,n+1}^{u,u}   - U_{n+1,n}^{u,u} \mp U_{n,n-1}^{l,l}     \mp U_{n-1,n}^{l,l},   \nonumber \\
\tilde{T}_n^{\pm}&=&  + U_{n,n+1}^{u,u}  - U_{n+1,n}^{u,u} \mp U_{n,n-1}^{l,l}     \pm U_{n-1,n}^{l,l},   \nonumber \\
\tilde{A}_n^{\pm}&=&  - U_{n,n}^{u,l}    + U_{n,n}^{l,u}   \pm U_{n+1,n-1}^{u,l}   \mp U_{n-1,n+1}^{l,u},  \nonumber \\
\tilde{B}_n^{\pm}&=&  - U_{n,n}^{u,l}    - U_{n,n}^{l,u}   \mp U_{n+1,n-1}^{u,l}   \mp U_{n-1,n+1}^{l,u}.
\end{eqnarray}

If we are interested in the electric current, we should add the velocities of the two inequivalent points of the BZ.
As a consequence, the $x$ component of the velocity vanishes while the $y$ component nearly doubles.

\end{document}